\documentclass[conference]{IEEEtran}

\usepackage{bm}
\usepackage{mathrsfs}
\usepackage{caption}
\usepackage{amsmath}
\allowdisplaybreaks[4]
\usepackage{lipsum}
\usepackage{mathtools}
\usepackage{subfigure}
\usepackage{amsmath} 
\usepackage{amssymb}
\usepackage{amsfonts}
\usepackage{verbatim} 
\usepackage{bm,color,soul}
\usepackage{algorithm}
\usepackage{algorithmic} 
\usepackage{cite}
\usepackage{caption}
\usepackage{stfloats}
\makeatletter
\renewcommand{\maketag@@@}[1]{\hbox{\m@th\normalsize\normalfont#1}}%
\makeatother
\newtheorem{remark}{Remark}
\usepackage[colorlinks=black, linkcolor=black, anchorcolor=black, citecolor=black]{hyperref} 
\hypersetup{hidelinks}

\usepackage{makecell}

\ifCLASSINFOpdf
\else
\fi
\usepackage{array}
\usepackage{balance}

\IEEEoverridecommandlockouts

\begin{document}
	%
	
	\title{Training-Free Energy Beamforming Assisted by Wireless Sensing}
	\author{
		\IEEEauthorblockN{Li Zhang$^{1}$, Yuan Fang$^{2,3}$, Zixiang Ren$^{1,2}$, Ling Qiu$^{1}$, and Jie Xu$^{3,2}$}
		\IEEEauthorblockA{$^1$Key Laboratory of Wireless-Optical Communications, Chinese Academy of Sciences,  \\
			School of Information Science and Technology,
			University of Science and Technology of China}
		\IEEEauthorblockA{$^2$Future Network of the Intelligence Institute (FNii), The Chinese University of Hong Kong (Shenzhen)}
		\IEEEauthorblockA{$^3$School of Science and Engineering (SSE), The Chinese University of Hong Kong (Shenzhen)}
		\IEEEauthorblockA
		{E-mails: {lzhang0228}@mail.ustc.edu.cn,
			{fangyuan}@cuhk.edu.cn, {rzx66}@mail.ustc.edu.cn,\\	{lqiu}@ustc.edu.cn, {xujie}@cuhk.edu.cn}\vspace{-1cm}  
		
		\thanks{L. Qiu and J. Xu are the corresponding authors.}
		
	}	
	\maketitle
	\vspace{-0.2cm}	
	
	\begin{abstract}
		This paper studies the transmit energy beamforming in a multi-antenna wireless power transfer (WPT) system, in which an access point (AP) equipped with a uniform linear array (ULA) sends radio signals to wirelessly charge multiple single-antenna energy receivers (ERs). Different from conventional energy beamforming designs that require the AP to acquire the channel state information (CSI) via training and feedback, we propose a new training-free energy beamforming approach assisted by wireless radar sensing, which is implemented based on the following two-stage protocol. In the first stage, the AP performs wireless radar sensing to estimate the path gain and angle parameters of the ERs for constructing the corresponding CSI. In the second stage, the AP implements the transmit energy beamforming based on the constructed CSI to efficiently charge these ERs in a fair manner. Under this setup, first, we jointly optimize the sensing beamformers and duration in the first stage to minimize the sensing duration, while ensuring a given   accuracy threshold for parameters estimation subject to the maximum transmit power constraint at the AP. Next, we optimize the energy beamformers in the second stage to maximize the minimum harvested energy by all ERs. In this approach,  the estimation accuracy threshold for the first stage is properly designed to balance the resource allocation between the two stages for optimizing the ultimate energy harvesting performance. Finally, numerical  results show that the proposed training-free energy beamforming design performs close to the  performance upper bound with perfect CSI, and outperforms the benchmark schemes without such joint optimization and that with isotropic transmission. 
	\end{abstract}
	\section{Introduction}
	Future sixth-generation (6G) wireless networks are expected to support massive Internet-of-things (IoT) devices to enable emerging applications such as smart home and smart city  \cite{1}. In practice, IoT devices are with small size, thus making the conventional battery-based energy supply unreliable. As a result, how to provide sustainable power supply for a large number of IoT devices in a cost-effective manner is becoming an important but challenging task for the success of IoT applications. Wireless power transfer (WPT) has emerged as a promising technology to resolve this issue for realizing sustainable zero-power IoT networks\cite{2}, in which base stations (BSs) and access points (APs) can be utilized for wirelessly charging IoT devices as energy receivers (ERs).

	Among various approaches, the multi-antenna transmit energy beamforming has been widely recognized as a promising WPT technique to enhance the energy transfer efficiency, in which multiple transmit antennas are deployed at the AP, such that the transmit signal beams can be steered towards the desired directions of the ERs. The implementation of energy beamforming highly relies on the availability of channel state information (CSI) at the AP. Conventionally, there have been three  approaches for the AP to acquire the forward-link CSI with the ERs based on channel training and feedback (e.g., \cite{2,3,4}). In the first approach, the ERs send pilots to the AP in the reverse link using the same frequency band as the forward link, and based on the pilots the AP can estimate the reverse-link channel, which is then utilized as the forward-link CSI by exploiting the channel reciprocity\cite{2}. In the second approach, the AP sends pilots in the forward link, based on which each ER estimates its associated CSI and then feeds it back to the AP after proper quantization and compression  \cite{3}. In the third approach, the AP adaptively adjusts its pilots over time in the forward link, and each ER measures the harvested energy levels over time and feeds them back to the AP, which then estimates the CSI based on the feedback energy measurements  \cite{4}. However, the above three conventional approaches require the ERs to perform baseband signal processing and/or active signal feedback transmission, which are energy-consuming and thus may seriously reduce the net harvested energy by the ERs. Therefore, it is of great importance to find new energy beamforming designs with light or even zero training/feedback. 
	
	Integrating wireless radar sensing in wireless networks has attracted growing interests in both academia and industry to enable integrated sensing and communications (ISAC) for 6G  \cite{5}. The radar sensing capability has shown its potential in facilitating the wireless communications. In particular, the BS and AP can use the radar sensing to estimate the channel parameters (e.g., angles) of communication users as sensing targets, and accordingly construct the communication CSI with reduced signaling overhead. For instance, the author in \cite{6,7} studied a radar-assisted predictive beamforming design for vehicle-to-infrastructure (V2I) communications, in which the echo signals reflected by the vehicles are exploited to track and localize them for facilitating the information beamforming. Motivated by the success of ISAC and sensing-assisted communications, we expect that the radar sensing can also be integrated into WPT systems as an effective solution to facilitate the transmit energy beamforming. In the literature, there has been prior work \cite{8} studying the integration of WPT and ISAC in a multi-functional wireless systems, in which the transmit beamforming at the hybrid AP is optimized to balance the performance tradeoffs among powering, sensing, and communication. However, how to exploit radar sensing to enable zero- or light-training energy beamforming has not been well investigated.
	
	In particular, this paper studies a sensing-assisted training-free energy beamforming approach in a multi-antenna WPT system, in which a multi-antenna AP equipped with a uniform linear array (ULA) sends radio signals to wirelessly charge multiple single-antenna ERs. By assuming the line-of-sight (LoS) channel between the AP and all ERs, we propose to utilize the radar sensing at the AP to acquire the CSI of the ERs for implementing energy beamforming. Towards this end, we present a two-stage transmission protocol, in which the transmission block of interest is divided into two stages for radar sensing and energy transmission, respectively. In the first stage, the AP sends radar sensing signals and collects the echo signals from the ERs to estimate their path gain and angle parameters, which are then used for constructing the corresponding CSI. To achieve optimized target estimation performance in this stage, the AP properly designs the sensing  beamforming based on the ERs' parameters that are  {\it a priori} known from the estimation in the previous block. In the second stage, the AP implements the transmit energy beamforming based on the constructed CSI in the first stage, such that the wireless energy is fairly delivered to the multiple ERs. Under this setup, first, we jointly optimize the sensing beamformers and duration in the first stage to minimize the sensing duration, while ensuring a given accuracy threshold for parameters estimation subject to the maximum transmit power constraint at the AP. Next, we optimize the energy beamformers in the second stage to maximize the minimum harvested energy by all ERs. In the proposed design, the estimation accuracy threshold for the first stage should be properly designed to balance the resource allocation between the two stages for optimizing the ultimate energy harvesting performance. Finally, numerical results show that the proposed training-free energy beamforming design performs close to the  performance upper bound with perfect CSI, and outperforms the benchmark schemes without such joint optimization and that with isotropic transmission. 
	
	\emph{Notations:} Matrices are denoted by bold uppercase letters, and vectors are represented by bold lowercase letters. For a square matrix $\mathbf{A}$, $\operatorname{tr}\left(\mathbf{A}\right)$ denotes
	its trace, and $\mathbf{A} \succeq \mathbf{0}$ means that $\mathbf{A}$ is positive semi-definite. For a vector $\mathbf{b}$, $\operatorname{diag}(\mathbf{b})$ denotes a diagonal matrix with $\mathbf{b}$ being its diagonal elements. For an arbitrary-size matrix $\mathbf{B}$, $\operatorname{rank}\left(\mathbf{B}\right)$, $\mathbf{B}^{T}$, $\mathbf{B}^{H}$, and $\mathbf{B}^{*}$ denote
	its rank, transpose, conjugate transpose, and 
	conjugate, respectively. For a complex
	number $a$, $|a|$ denotes its magnitude, and $\text{arg}(a)$ denotes its phase. $\mathbb{E}(\cdot)$
	denotes the stochastic expectation, and $\|\cdot\|$ denotes the
	Euclidean norm of a vector. $\mathbb{C}^{M \times N}$ denotes
	the space of $M \times N$ complex matrices. $\mathbf{A}\odot\mathbf{B}$ represents the Hadamard
	product of two matrices $\mathbf{A}$ and $\mathbf{B}$. Furthermore, we denote $j=\sqrt{-1}$.

	\section{System Model}

	\begin{figure}
		\setlength{\abovecaptionskip}{-10pt}
		\setlength{\belowcaptionskip}{-15pt}
		\centering
			\centering
			\includegraphics[width= 0.45\textwidth]{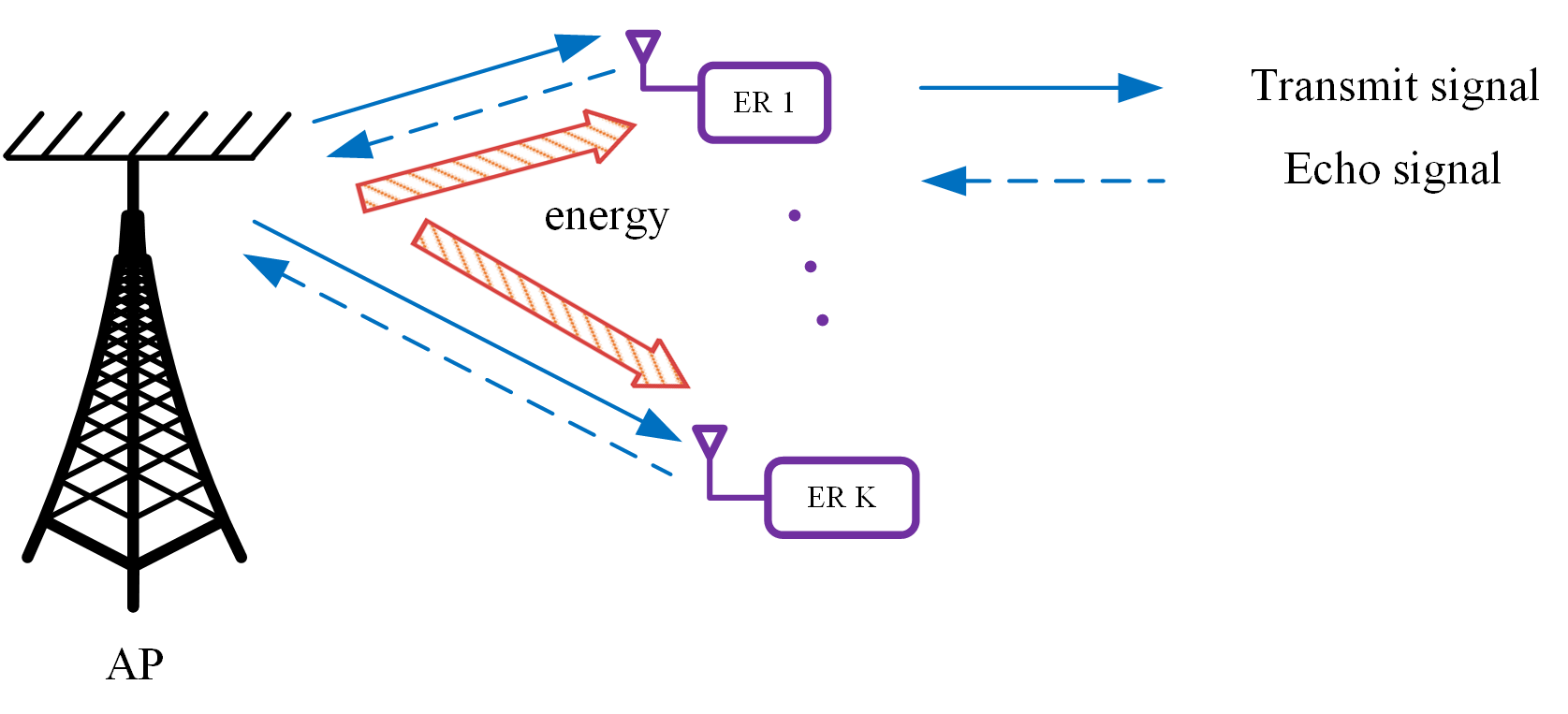}
		\DeclareGraphicsExtensions.
		\captionsetup{font={scriptsize},skip=3pt}
		\caption{Illustration of a sensing-assisted WPT system.}
		\label{figure1}
	\end{figure}
	As shown in Fig. 1, we consider a  multi-antenna WPT system, in which a multi-antenna AP with a ULA transmits wireless energy to $K$ single-antenna ERs. The AP is equipped with $N_t$ transmit antennas and $N_r$ receive antennas. Let $\mathcal{K} \triangleq\{1, \ldots, K\}$ denote the set of ERs. Each ER is equipped with a radio frequency (RF) energy harvesting module for harvesting energy from the AP. It is assumed that $K \leq N_t \leq N_r$.
	
	As shown in Fig. 2, we consider the block-based transmission by assuming quasi-static channel models. Let $T$ denote the duration of each particular transmission block in the number of symbols. It is assumed that the ERs' locations and wireless channels remain unchanged over each block but may change over different blocks due to the mobility of ERs. Let $\theta_{k}$ and $d_{k}$ denote the angle and distance between the AP and each ER $k$ in the current block of interest, and $\bar \theta_{k}= \theta_{k} + \Delta \theta_{k}$ and $\bar d_{k}=d_{k} +\Delta d_{k}$ denote the estimated angle and distance in the previous
	block, where $\Delta \theta_{k}$ and  $\Delta d_{k}$ denote the corresponding estimation errors. It is assumed that $\Delta \theta_{k}$ and $\Delta d_{k}$ are random variables that are bounded, i.e., $|\Delta \theta_{k}| \leq \phi$ and $|\Delta d_{k}|\leq D$, with $\phi$ and $D$ denoting the corresponding error bounds. As $\bar\theta_{k}$ and $\bar d_{k}$ are known prior to transmission in each block, it follows that $\theta_{k}$ and $d_{k}$ are random variables lying in the regions $\left[\bar{\theta}_{k}-\phi,\bar{\theta}_{k}+\phi\right]$ and $\left[\bar{d}_{k}-D,\bar{d}_{k}+D\right]$, respectively.
	
	We consider a two-stage transmission protocol, in which the transmission block with duration $T$ is divided into two stages with $\tau$ and $T-\tau$ symbols for radar sensing and energy transmission, respectively. Here, $\tau$ is a design parameter to be determined later. In the first stage, the AP performs radar sensing to estimate the path gain and angle parameters of the ERs in the current block of interest for constructing the corresponding CSI, in which the radar sensing beamforming can be designed based on the estimations in the previous block. In the second stage, the AP implements the transmit energy beamforming for fairly charging multiple ERs, in which the beamformers are designed based on the constructed CSI. In order to facilitate the sensing in the first stage, it is assumed that the ERs do not harvest energy in this stage by properly adjusting their antenna impedance. 
	
	We consider the LoS channel model between the AP and ERs, since the WPT is implemented in a short distance such that the LoS channel normally dominates the non-LoS (NLoS) paths. Let $\mathbf{h}_{k}\in \mathbb{C}^{N_{t} \times 1}$ denote the LoS channel from the AP to each ER $k$. In particular,  $\mathbf{h}_{k}$ is given by \cite{6} \begin{equation}\label{1}
		\setlength\abovedisplayskip{4pt}		
		\setlength\belowdisplayskip{4pt}
		\mathbf{h}_{k}=\sqrt{\rho_{0} d_{k}^{-2}} e^{j \frac{2 \pi}{\lambda} d_{k}}\mathbf{a}_{t}\left(\theta_{k}\right),
	\end{equation}
	where $\rho_{0}$ denotes the channel power at reference distance $d_{0}=1 $ m, $d_{k}$ denotes the distance between the AP to ER $k$, $\lambda$ denotes the carrier wavelength, $\theta_{k}$ is the direction of arrival (DoA) of ER $k$ relative to the AP, and $	\mathbf{a}_{t}\left(\theta_{k}\right)$ denotes the steering vector at the transmitter of AP, i.e.,
	\begin{equation}\label{transmit vector}
		\setlength\abovedisplayskip{4pt}		
		\setlength\belowdisplayskip{4pt}
		\mathbf{a}_{t}\left(\theta_{k}\right)=\left[1,e^{j 2\pi \frac{\tilde{d}}{\lambda} \sin \theta_{k}}, \ldots, e^{j  2\pi \left(N_{t}-1\right) \frac{\tilde{d}}{\lambda} \sin \theta_{k}}\right]^{T},
	\end{equation}
	where $\tilde{d}$ denotes the spacing between adjacent antennas.
	\vspace{-0.1cm}
	\subsection{Radar Sensing Stage}
	First, we consider the MIMO radar sensing over the coherence processing interval (CPI) with $\tau \geq N_{t}$ symbols in the first stage. Let $\mathbf{x}(t) \in \mathbb{C}^{N_{t} \times 1}$ denote the transmitted signal for sensing at symbol $t$, where $t\in\{1,\ldots, \tau\}$, and $\mathbf{X}=\left[\mathbf{x}(1), \ldots, \mathbf{x}(\tau)\right] \in \mathbb{C}^{N_{t} \times \tau}$ denote the transmitted signals over the CPI. The sample covariance
	matrix is given by $\mathbf{S}_{x}=\frac{1}{\tau}\mathbf{X}\mathbf{X}^{H} \succeq \mathbf{0}$, which is an optimization variable to
	be designed in the first stage. Let $P_{\text{max}}$ denote the maximum transmit power at the AP, and then we have the transmit power constraint as
	\vspace{-0.05cm}
	\begin{equation}
		\setlength\abovedisplayskip{4pt}		
		\setlength\belowdisplayskip{4pt}
		\operatorname{tr}\left(\mathbf{S}_{x}\right)\leq P_{\text{max}}. 
	\end{equation}
	
	The received echo signal $\mathbf{y}(t)\in \mathbb{C}^{N_{\mathrm{r}} \times 1}$ by the AP from the ERs at symbol $t$  is given by 
	\begin{equation}\label{receive signal t}
		\setlength\abovedisplayskip{4pt}		
		\setlength\belowdisplayskip{4pt}
		\mathbf{y}(t)=\sum_{k=1}^{K} \alpha_{k} \mathbf{a}_{r}\left(\theta_{k}\right) \mathbf{a}_{t}^{T}\left(\theta_{k}\right)  \mathbf{x}(t) + \mathbf{z}(t), t\in\{1,\ldots, \tau\},
	\end{equation}
	where $\alpha_{k} \in \mathbb{C}$ represents the complex path gain between the AP and ER $k$ accounting for the round-trip path-loss and the radar cross section (RCS) of ER $k$, $\mathbf{z}(t)$ denotes the additive white Gaussian noise (AWGN) with mean zero and
	covariance $\sigma_{r}^{2}\mathbf{I}$, and  $\mathbf{a}_{r}\left(\theta_{k}\right) \in \mathbb{C}^{N_{\mathrm{r}} \times 1}$ denotes the steering vector at the receiver of AP, i.e.,
	\begin{equation}\label{transmit vector}
		\setlength\abovedisplayskip{4pt}		
		\setlength\belowdisplayskip{4pt}
		\mathbf{a}_{r}\left(\theta_{k}\right)=\left[1,e^{j 2\pi \frac{\tilde{d}}{\lambda} \sin \theta_{k}}, \ldots, e^{j  2\pi \left(N_{r}-1\right) \frac{\tilde{d}}{\lambda} \sin \theta_{k}}\right]^{T}.
	\end{equation}
	By defining
	$\mathbf{Y}=\left[\mathbf{y}(1), \ldots, \mathbf{y}(\tau)\right] \in \mathbb{C}^{N_{\mathrm{r}} \times \tau}$ and $\mathbf{Z}=\left[\mathbf{z}(1), \ldots, \mathbf{z}(\tau)\right] \in \mathbb{C}^{N_{\mathrm{r}} \times \tau}$, the received echo signal $\mathbf{y}(t)$ in \eqref{receive signal t} over the CPI at the AP is rewritten as
	\begin{equation}\label{receive signal}
		\setlength\abovedisplayskip{4pt}		
		\setlength\belowdisplayskip{4pt}
		\mathbf{Y}=\sum_{k=1}^{K} \alpha_{k} \mathbf{a}_{r}\left(\theta_{k}\right) \mathbf{a}_{t}^{T}\left(\theta_{k}\right)  \mathbf{X} + \mathbf{Z}.
	\end{equation}  
	Based on the received echo signal $\mathbf{Y}$ in this stage, the AP estimates angle $\theta_{k}$ and round-trip path gain $\alpha_{k}$ of ER $k$ in the current block of interest by parameter
	estimation techniques such as Capon and approximate maximum likelihood (CAML) algorithms\cite{9}. Let $\hat{\theta}_{k}$ and $\hat{\alpha}_{k}$ denote the estimated angle and path gain of ER $k$, which are then used for constructing the corresponding CSI of ER $k$ for energy transmission. 
	\subsection{Energy Transmission Stage}
	Next, we consider the energy transmission in the second stage. Let $\mathbf{x}(t) \in \mathbb{C}^{N_{t} \times 1}$ denote the transmitted signal for energy transmission at symbol $t$, where $t\in\{\tau+1,\ldots, T\}$, and $\mathbf{R}_{x}=\mathbb{E}\left\{\mathbf{x}(t) \mathbf{x}^{H}(t)\right\} \succeq \mathbf{0}$ denote the transmit energy covariance matrix in this stage. Note that for our considered transmit energy beamforming, $\mathbf{R}_{x}$ is assumed to be of general rank,
	i.e., $m=\mathrm{rank}(\mathbf{R}_{x}) \leq N_{t}$. This corresponds to the case
	with $m$ energy beams, each of which can be obtained via
	the eigenvalue decomposition (EVD) of $\mathbf{R}_{x}$. The received signal at ER $k$ is expressed as follows by ignoring the receiver noise that is practically negligible for ERs 
	\begin{equation}\label{rece1}
		\setlength\abovedisplayskip{4pt}		
		\setlength\belowdisplayskip{4pt}
		y_{k}(t)=\mathbf{h}_{k}^{H} \mathbf{x}(t)= \sqrt{\rho_{0} d_{k}^{-2}} e^{j \frac{2 \pi}{\lambda} d_{k}}\mathbf{a}_{t}\left(\theta_{k}\right)\mathbf{x}(t). 
	\end{equation}
	Due to the broadcast nature of wireless channels, the energy carried by all energy beams can be
	harvested at each ER. As a result, the received RF power (energy over a unit time period, in Watt) at ER $k$ is \footnote{Notice that ER $k$ exploits the rectifiers to convert the received RF signals into direct current (DC) signals for energy harvesting. Here, we consider the received RF power  as the energy harvesting performance metric, as the harvested DC power is in general monotonically non-decreasing with respect to the received RF power.}
	\begin{equation}
		\setlength\abovedisplayskip{4pt}		
		\setlength\belowdisplayskip{4pt}
		\begin{aligned}
			P_{k}(\mathbf{h}_{k},\mathbf{R}_{x}) &= \left|\mathbf{h}_{k}^{H} \mathbf{x}(t)\right|^{2}  \\
			&= \rho_{0} d_{k}^{-2}  \operatorname{tr}\left(\mathbf{a}_{t}^{H}\left(\theta_{k}\right) \mathbf{R}_{x} \mathbf{a}_{t}\left(\theta_{k}\right)\right).  
		\end{aligned}
	\end{equation}
	
	To design the transmit energy covariance matrix $\mathbf{R}_{x}$ for fairly charging multiple ERs, we need to know the CSI from the AP to each ER $k$, i.e., $\mathbf{h}_{k}$. The corresponding CSI can be constructed based on the estimated angle $\hat{\theta}_{k}$ and path gain $\hat{\alpha}_{k}$ of ER $k$ in the first stage. According to the components of the path gain $\alpha_{k}$, the modulus of path gain is given by \cite{10}\begin{equation}\label{alpha}
		\setlength\abovedisplayskip{4pt}		
		\setlength\belowdisplayskip{4pt}
		\left|\alpha_{k}\right|=\sqrt{\rho_{0}  d_{k}^{-2} \times \beta_{k}  d_{k}^{-2}},
	\end{equation} 
	\begin{figure}
		\setlength{\abovecaptionskip}{-10pt}
		\setlength{\belowcaptionskip}{-15pt}
		\centering
			\centering
			\includegraphics[width= 0.37\textwidth]{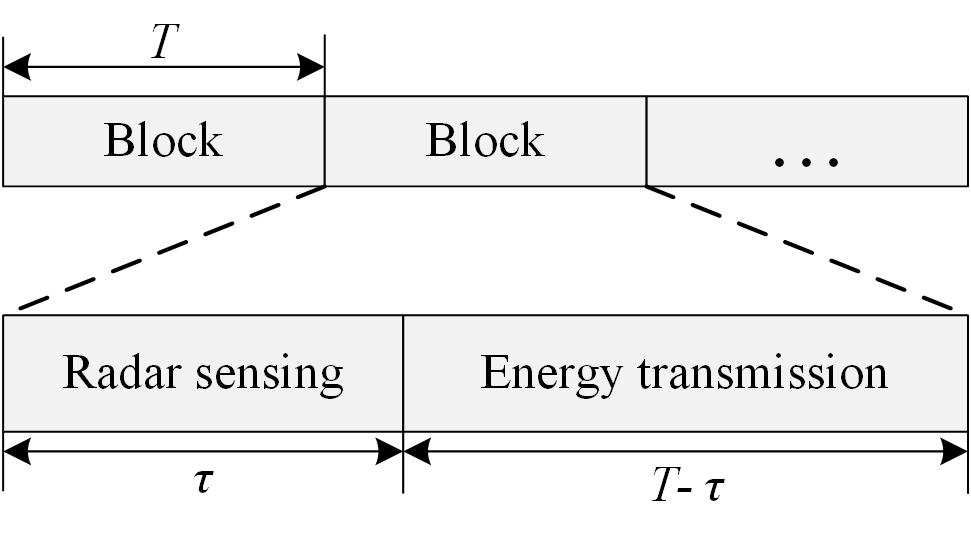}
		\DeclareGraphicsExtensions.
		\captionsetup{font={scriptsize},skip=3pt}
		\caption{Block-based transmission with two-stage protocol.}
		\label{figure2}%
	\end{figure}%
	where $\beta_{k}$ denotes the RCS of ER $k$. We assume that the AP can infer $\beta_{k}$ based on the long-term estimation. Besides, the phase of path gain $\alpha_{k}$ is $\frac{2 \pi}{\lambda} 2d_{k}$, whose value is twice the phase of the CSI $\mathbf{h}_{k}$ in $\eqref{1}$. Consequently, based on the components of the CSI $\mathbf{h}_{k}$, the constructed CSI of ER $k$ by using the estimated angle $\hat{\alpha}_k$ and path gain $\hat{\theta}_k$ can be denoted as
	\begin{equation}\label{h}
		\setlength\abovedisplayskip{4pt}		
		\setlength\belowdisplayskip{4pt}
		\hat{\mathbf{h}}_{k}=\sqrt{\left|\hat{\alpha}_{k}\right|} \tilde{\beta}_{k}   e^{j \frac{\text{arg}(\bar{\alpha}_{k})}{2}}\mathbf{a}_{t}\left(\hat{\theta}_{k}\right),
	\end{equation}
	where $\tilde{\beta}_{k}=\left(\rho_{0}\beta_{k}^{-1}\right)^{\frac{1}{4}}$ is a constant known to the AP. Correspondingly, the AP can utilize the constructed $\hat{\mathbf{h}}_k$ to optimize $\mathbf{R}_{x}$ for maximizing $P_{k}(\hat{\mathbf{h}}_{k},\mathbf{R}_{x})$. 
	\vspace{-0.2cm}
	\section{Proposed Sensing Duration and Transmit Beamforming Design}
	\vspace{-0.1cm}
	In this section, we propose to optimize the sensing/energy beamforming in the two stages and the sensing duration to improve the energy harvesting performance. Specifically, in the first stage, we properly design the sensing duration and beamforming based on the ERs' angles and path gains that are {\it a priori} known from the estimation in the previous block, in which a predetermined estimation accuracy threshold $\Gamma$ is ensured. In the second stage, based on the constructed CSI $\hat{\mathbf{h}}_{k}$ in the first stage, we design the energy beamforming to maximize the minimum harvested power among all ERs. 
	\vspace{-0.1cm}
	\subsection{Sensing Duration and Beamforming Design in the First Stage} 
	\vspace{-0.1cm}
	In this stage, we focus on the sensing task, aiming to estimate the angles and path gains of the ERs in the current block of interest. To optimize the target estimation performance, the AP designs the sensing duration $\tau$ and sample covariance matrix $\mathbf{S}_{x}$ based on the known angles and path gains estimated in the previous block. Our objective is to minimize the sensing duration while ensuring the estimation accuracy requirement, such that the minimum transmission energy is utilized for the first stage. In particular, we use the Cramér-Rao bound (CRB) as the performance metric for parameters estimation, which serves as a performance
	lower bound for any unbiased estimator.
	
	In particular, the CRB matrix for estimating angles and path gains of the ERs is given as follows. First, the angles and path gains of all ERs to be estimated are denoted as $\boldsymbol{\theta} =\left[\theta_{1}, \cdots, \theta_{k},\cdots, \theta_{K}\right]^{T}$ and
	$\mathbf{b}=\left[\alpha_{1}, \cdots,\alpha_{k},\cdots, \alpha_{K}\right]^{T}$.
	
	For notational convenience, the receive signal in \eqref{receive signal} is re-arranged as
	\vspace{-0.1cm}
	\begin{equation}
		\setlength\abovedisplayskip{4pt}		
		\setlength\belowdisplayskip{4pt}
		\mathbf{Y}=\mathbf{A}_{r} \mathbf{B A}_{t}^{T} \mathbf{X}+\mathbf{Z},
	\end{equation}
	where $\mathbf{A}_{r}  =\left[\mathbf{a}_{r}\left(\theta_{1}\right), \cdots, \mathbf{a}_{r}\left(\theta_{K}\right)\right]$, $\mathbf{B} =\operatorname{diag}(\mathbf{b})$, and $\mathbf{A}_{t} =\left[\mathbf{a}_{t}\left(\theta_{1}\right), \cdots, \mathbf{a}_{t}\left(\theta_{K}\right)\right]$.

	The Fisher information matrix (FIM) with respect to $\boldsymbol{\theta}$, and the real
	and imaginary parts of $\mathbf{b}$ is given by \cite{11}
	\begin{equation}\label{F}
		\setlength\abovedisplayskip{4pt}		
		\setlength\belowdisplayskip{4pt}
		\mathbf{F}=\frac{2}{\sigma_{r}^{2}}\left[\begin{array}{ccc}
			\operatorname{Re}\left(\mathbf{F}_{11}\right) & \operatorname{Re}\left(\mathbf{F}_{12}\right) & -\operatorname{Im}\left(\mathbf{F}_{12}\right) \\
			\operatorname{Re}^{T}\left(\mathbf{F}_{12}\right) & \operatorname{Re}\left(\mathbf{F}_{22}\right) & -\operatorname{Im}\left(\mathbf{F}_{22}\right) \\
			-\operatorname{Im}^{T}\left(\mathbf{F}_{12}\right) & -\operatorname{Im}^{T}\left(\mathbf{F}_{22}\right) & \operatorname{Re}\left(\mathbf{F}_{22}\right)
		\end{array}\right],
	\end{equation}
	where $\mathbf{F}_{11}= \tau\left(\dot{\mathbf{A}}_{r}^{H} \dot{\mathbf{A}}_{r}\right) \odot\left(\mathbf{B}^{H} \mathbf{A}_{t}^{H} \mathbf{S}_{x}^{*} \mathbf{A}_{t} \mathbf{B}\right)+\tau\left(\dot{\mathbf{A}}_{r}^{H} \mathbf{A}_{r}\right) \\\odot \left(\mathbf{B}^{H} \mathbf{A}_{t}^{H} \mathbf{S}_{x}^{*} \dot{\mathbf{A}}_{t} \mathbf{B}\right) +\tau\left(\mathbf{A}_{r}^{H} \dot{\mathbf{A}}_{r}\right) \odot\left(\mathbf{B}^{H} \dot{\mathbf{A}}_{t}^{H} \mathbf{S}_{x}^{*} \mathbf{A}_{t} \mathbf{B}\right)+\tau\left(\mathbf{A}_{r}^{H} \mathbf{A}_{r}\right) \odot\left(\mathbf{B}^{H} \dot{\mathbf{A}}_{t}^{H} \mathbf{S}_{x}^{*} \dot{\mathbf{A}}_{t} \mathbf{B}\right)$, $\mathbf{F}_{12}=\tau\left(\dot{\mathbf{A}}_{r}^{H} \mathbf{A}_{r}\right) \odot\left(\mathbf{B}^{H} \mathbf{A}_{t}^{H} \mathbf{S}_{x}^{*} \mathbf{A}_{t}\right)+\tau\left(\mathbf{A}_{r}^{H} \mathbf{A}_{r}\right) \odot\left(\mathbf{B}^{H} \dot{\mathbf{A}}_{t}^{H} \mathbf{S}_{x}^{*} \mathbf{A}_{t}\right)$,  $\mathbf{F}_{22}=\tau\left(\mathbf{A}_{r}^{H} \mathbf{A}_{r}\right) \odot\left(\mathbf{A}_{t}^{H} \mathbf{S}_{x}^{*} \mathbf{A}_{t}\right)$, $\dot{\mathbf{A}}_{r}=\left[\begin{array}{lll}
		\dot{\mathbf{a}}_{r}(\theta_{1}), \cdots,  \dot{\mathbf{a}}_{r}(\theta_{K})
	\end{array}\right]$, and $\dot{\mathbf{A}}_{t}=\left[\begin{array}{lll}
	\dot{\mathbf{a}}_{t}(\theta_{1}), \cdots, \dot{\mathbf{a}}_{t}(\theta_{K})
	\end{array}\right]$.
	The derivative of $\mathbf{a}_t(\theta_k)$ is
	\begin{equation}\label{deri trans vec}
		\setlength\abovedisplayskip{2pt}		
		\setlength\belowdisplayskip{2pt}
		\dot{\mathbf{a}}_{t}(\theta_{k})=\left[0,j  2\pi \frac{\tilde{d}}{\lambda} a_{2}\cos \theta_{k}, \ldots, j  2\pi\left(N_{t}-1\right) \frac{\tilde{d}}{\lambda} a_{N_{t}}  \cos \theta_{k}\right]^{T},
	\end{equation}		
	where $a_{i}$ represents the $i \text {-th}$ entry of $\mathbf{a}_{t}(\theta_{k})$. And the derivative of $\mathbf{a}_{r}(\theta_{k})$ takes similar form to \eqref{deri trans vec}.	
	Then, based on the FIM in \eqref{F}, the corresponding CRB matrix
	is $\mathbf{C}=\mathbf{F}^{-1}$. In particular, we adopt the trace of the CRB matrix $\mathbf{C}$ as the performance
	metric to be optimized, which has been shown to be a good design choice to lower the CRB of multiple targets \cite{11}. Accordingly, the CRB for estimating parameters $\boldsymbol{\theta}$ and $\mathbf{b}$ under a given duration $\tau$ and sample covariance matrix $\mathbf{S}_{x}$  is denoted as
	\begin{equation}\label{CRB}
		\setlength\abovedisplayskip{4pt}		
		\setlength\belowdisplayskip{4pt}
		\text{CRB}\left(\tau,\mathbf{S}_{x},\boldsymbol{\theta},\mathbf{b}\right) =\operatorname{tr}\left(\mathbf{C}\right)= \operatorname{tr}\left(\mathbf{F}^{-1}\right).
	\end{equation}
	Notice that the CRB in \eqref{CRB} depends on the parameters $\boldsymbol{\theta}$ and $\mathbf{b}$, which are generally unknown initially. In the following, we optimize the estimation performance based on the estimated  $\bar{\boldsymbol{\theta}}=\left[\bar{\theta}_{1}, \cdots, \bar{\theta}_{k},\cdots, \bar{\theta}_{K}\right]^{T}$ and $\bar{\mathbf{b}}=\left[\bar{\alpha}_{1}, \cdots,\bar{\alpha}_{k},\cdots, \bar{\alpha}_{K}\right]^{T}$  in the previous block, and accordingly  use 
	$\text{CRB}\left(\tau,\mathbf{S}_{x},\bar{\boldsymbol{\theta}},\bar{\mathbf{b}}\right)$ as an apprixmation of 
	$\text{CRB}\left(\tau,\mathbf{S}_{x},\boldsymbol{\theta},\mathbf{b}\right)$.
	
	Furthermore, note that allocating more duration for sensing can ensure the estimation accuracy requirement, but it may also lead to a decrease in the available duration for energy beamforming in the second stage. Therefore, in the first stage, we aim to jointly optimize sensing duration $\tau$ and sample covariance matrix $\mathbf{S}_{x}$ to minimize sensing duration $\tau$ while ensuring the estimation accuracy characterized by the threshold $\Gamma$, subject to the maximum transmit power constraint at the AP.
	As such, the corresponding CRB constrained sensing duration minimization problem is formulated as
	\vspace{-0.2cm}
	\begin{subequations}
		\begin{align}
			\text{(P1)}:	\min _{\tau,\mathbf{S}_{x}\succeq 0} 
			&\quad \tau\\ 
			\mathrm{s.t.}
			&\quad \tau \in \{N_{t}, \ldots, T\},\label{22b}\\			 
			&\quad\operatorname{tr}\left(\mathbf{S}_{x}\right)\leq P_{\text{max}}, \label{C2}\\
			&\quad \text{CRB}\left(\tau,\mathbf{S}_{x},\bar{\boldsymbol{\theta}},\bar{\mathbf{b}}\right) \leq \Gamma.\label{C3}
		\end{align}
	\end{subequations}
	Problem (P1) is non-convex, as constraint \eqref{C3} is non-convex due to the coupling between $\tau$ and $\mathbf{S}_{x}$. In the following, we present the optimal solution to problem (P1).

	 First, it is noted that the optimal solution to (P1) is attained when constraint \eqref{C2} is met with strict equality, since otherwise the AP can further increase the transmit power to reduce the objective value while achieving the same CRB in \eqref{C3}. Furthermore, as the objective value of (P1) is independent from $\mathbf{S}_{x}$, the optimal solution of $\mathbf{S}_{x}$ in (P1) can be obtained by minimizing $\text{CRB}\left(\tau,\mathbf{S}_{x},\bar{\boldsymbol{\theta}},\bar{\mathbf{b}}\right)$ under any given value of $\tau$. As such, the optimal solution of $\mathbf{S}_{x}$ can be obtained by solving the following problem:
	 \vspace{-0.2cm}
	\begin{subequations}
		\begin{align}		
			\setlength\abovedisplayskip{1pt}		
			\setlength\belowdisplayskip{1pt}
			\text{(P2)}:	\min _{\mathbf{S}_{x}\succeq 0} 
			&\quad \text{CRB}\left(\tau,\mathbf{S}_{x},\bar{\boldsymbol{\theta}},\bar{\mathbf{b}}\right)\\ 
			\mathrm{s.t.}		
			&\quad\operatorname{tr}\left(\mathbf{S}_{x}\right)= P_{\text{max}}. \label{Pmax}		
		\end{align}
	\end{subequations}To solve problem (P2), we introduce auxiliary variables $\left\{t_{i}\right\}_{i=1}^{3K}$. Then, the problem (P2) is reformulated as 
		\vspace{-0.2cm}
	\begin{subequations}
		\begin{align}
			\setlength\abovedisplayskip{1pt}		
			\setlength\belowdisplayskip{1pt}
			\text{(P2.1)}:	\min _{\mathbf{S}_{x}\succeq 0,\left\{t_{i}\right\}_{i=1}^{3K}} 
			& \sum_{i=1}^{3 K} t_{i} \\ 
			\mathrm{s.t.}\quad
			& \boldsymbol{e}_{i}^{T} \mathbf{F}^{-1} \boldsymbol{e}_{i} \leq t_{i}, \forall i \in \{1, \ldots, 3 K\}\label{24b}\\		
			& \eqref{Pmax} \notag,		
		\end{align}
	\end{subequations}
	where $\mathbf{e}_{i}$ denotes the $i \text{-th}$ column
	of identity matrix $\mathbf{I}_{3K}$. Further, the constraint in $\eqref{24b}$ can be equivalently transformed into linear matrix
	inequalities (LMIs) with respect to $\mathbf{S}_{x}$, i.e., 
	\begin{equation}
		\setlength\abovedisplayskip{4pt}		
		\setlength\belowdisplayskip{4pt}
		\left[\begin{array}{cc}\label{cc4}
			\setlength\abovedisplayskip{1pt}		
			\setlength\belowdisplayskip{1pt}
			\mathbf{F} & \boldsymbol{e}_{i} \\
			\boldsymbol{e}_{i}^{T} & t_{i}
		\end{array}\right] \succeq \mathbf{0}, \forall i \in \{1, \ldots, 3 K\}.
	\end{equation}
	Accordingly, the problem (P2.1) is reformulated as
	\begin{equation}
			\setlength\abovedisplayskip{4pt}		
		\setlength\belowdisplayskip{4pt}
		\begin{aligned}\label{P2.2}
			\setlength\abovedisplayskip{1pt}		
			\setlength\belowdisplayskip{1pt}
			(\text{P2.2}):	\min _{\mathbf{S}_{x}\succeq 0,\left\{t_{i}\right\}_{i=1}^{3K}} 
			&\sum_{i=1}^{3 K} t_{i}\\ 
			\mathrm{s.t.}\quad
			&\eqref{Pmax},~\text{and}~\eqref{cc4},\\
		\end{aligned}
	\end{equation}
	which is a semi-definite program (SDP) that can be optimally solved by convex solvers, such as CVX \cite{12}. Let $\mathbf{S}_{x}^{\star}$ denote the obtained optimal solution to problem (P2.2). 
	
	Next, with $\mathbf{S}_{x}^{\star}$ obtained, we find the optimal sensing duration $\tau$ by solving the following problem: 
	\vspace{-0.1cm}
	\begin{subequations}
		\begin{align}\label{P3}
				\setlength\abovedisplayskip{4pt}		
			\setlength\belowdisplayskip{4pt}
			\text{(P3)}:	\min _{\tau} 
			&\quad \tau\\ 
			\mathrm{s.t.}	
			&\quad\text{CRB}\left(\tau,\mathbf{S}_{x}^{\star},\bar{\boldsymbol{\theta}},\bar{\mathbf{b}}\right) \leq \Gamma,\label{27b}\\
			&\quad \eqref{22b} \notag.
		\end{align}
	\end{subequations}
	Recall that $\text{CRB}\left(\tau,\mathbf{S}_{x},\bar{\boldsymbol{\theta}},\bar{\mathbf{b}}\right)$ is inversely proportional to $\tau$. Therefore, the optimal solution of $\tau$ to (P3) is obtained as the minimum sensing duration $\tau^{\star}$ to ensure CRB threshold $\Gamma$. Thus, the optimal solutions of $\mathbf{S}_{x}^{\star}$ and $\tau^\star$ to problem (P1) are finally obtained.
	
	\subsection{Energy Beamfroming Design in the Second Stage}
	In this stage, we consider the energy transmission during the remaining duration of  $T-\tau^{\star}$. In particular, the transmit energy covariance matrix $\mathbf{R}_{x}$ is  designed to maximize the minimum harvested RF
	power among all the ERs based on constructed CSI $\hat{\mathbf{h}}_{k}$ as described in Section II-B.  In this case, the minimum harvested RF power maximization problem is formulated as
	\vspace{-0.2cm}
	\begin{subequations}
		\begin{align}\label{P2}
			\setlength\abovedisplayskip{4pt}		
			\setlength\belowdisplayskip{4pt}
			(\text{P4}):	\max _{\mathbf{R}_{x}\succeq 0} 
			&\quad \min _{k \in \mathcal{K}}   \hat{\boldsymbol{h}}_{k}^{H} \mathbf{R}_{x} \hat{\boldsymbol{h}}_{k}\\ 
			\mathrm{s.t.}
			& \quad\operatorname{tr}\left(\mathbf{R}_{x}\right)\leq P_{\text{max}}\label{C23}.
		\end{align}
	\end{subequations}
	By introducing an auxiliary variable $E$, the optimization  problem (\text{P4}) is simplified as 
	\vspace{-0.2cm}
	\begin{subequations}
		\begin{align}\label{P4}
			\setlength\abovedisplayskip{3pt}		
			\setlength\belowdisplayskip{3pt}
			(\text{P4.1}):	\max _{\mathbf{R}_{x},E} 
			&\quad E\\ 
			\mathrm{s.t.}
			&\quad \hat{\boldsymbol{h}}_{k}^{H} \mathbf{R}_{x} \hat{\boldsymbol{h}}_{k} \geq E, \forall k \in \mathcal{K}\\
			&\quad \eqref{C23} \notag, 
		\end{align}	
	\end{subequations}%
which is also a SDP that can be optimally solved by CVX. Let $\mathbf{R}_{x}^{\star}$ denote the optimal solution to
	(P4.1). Then, the corresponding minimum average harvested RF power among all the ERs is given by $\frac{(T-\tau^{\star}) }{T}\mathop{\min}\limits_{k\in\mathcal K} \left\{\boldsymbol{h}_{1}^{H} \mathbf{R}_{x}^{\star} \boldsymbol{h}_{1}, \ldots, \boldsymbol{h}_{K}^{H} \mathbf{R}_{x}^{\star} \boldsymbol{h}_{K} \right\}$, where $\boldsymbol{h}_{k}$ is the real channel between the AP and ER $k$.
	\begin{remark}\label{remark}
		Note that in our proposed design, the estimation accuracy threshold $\Gamma$ is a key parameter that can be designed to balance the resource allocation between the two stages for maximizing the ultimate energy harvesting performance. In particular, by reducing the CRB threshold $\Gamma$, the time duration allocated for radar sensing $\tau$ in the first stage should be increased to meet the more stringent estimation accuracy requirement. This leads to more accurate channel estimation in the first stage, but results in less duration for energy transmission in the second stage. This phenomenon will be demonstrated in the numerical results in Section IV.
	\end{remark} 
		\begin{figure}
		\setlength{\abovecaptionskip}{-10pt}
		\setlength{\belowcaptionskip}{-15pt}
		\centering
			\centering
			\includegraphics[width= 0.39\textwidth]{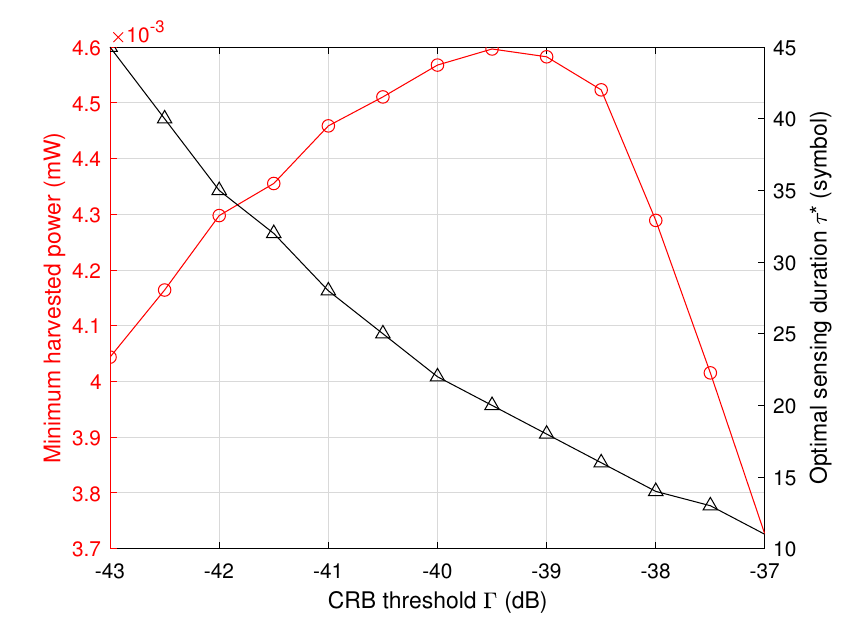}
		\DeclareGraphicsExtensions.
		\captionsetup{font={scriptsize},skip=3pt}
		\caption{The minimum harvested power and the optimized sensing duration $\tau^\star$ versus the CRB threshold $\Gamma$.}
		\label{figure3}
		\vspace{-0.1cm}
	\end{figure}	
	\vspace{-0.05cm}	
	\section{Numerical Results}
	\vspace{-0.05cm}	
	In this section, we present numerical results to validate the effectiveness of our proposed  sensing-assisted energy beamforming design. In the simulation, the interval
	between adjacent antennas of the AP is half-wavelength.  We set the noise power as $\sigma_{r}^{2}=-50$ dBm and the channel power at the reference distance as $\rho_{0}=-40$ dB. We consider that there are $K=3$ ERs in the system, in which their angles $\{\bar \theta_k\}$ are given by $0^{\circ}$, $30^{\circ}$, and $60^{\circ}$, and the distances $\{\bar{d}_k\}$ are $5$ m, $8$ m, and $10$ m, respectively. The estimation error bounds  are set as $\phi=5^{\circ}$, $D=2$ m, respectively. Furthermore, we assume that each transmission block consists of $T=200$ symbol periods.
	
	Fig.~\ref{figure3} shows the minimum average  harvested power among the ERs (i.e., energy normalized by the block duration $T$) versus the CRB threshold $\Gamma$ for the first stage, where the maximum transmit power is set to be $P_{\text{max}}=30$ dBm and the numbers of transmit and receive antennas at the AP are set as $N_{t}=N_{r}=8$. It is observed that there exists an optimal value of $\Gamma$ that achieves the maximum minimum harvested power. When $\Gamma$ is lower than this value, the minimum harvested power increases with the increase of $\Gamma$. This is due to the fact that the AP needs to allocate less sensing duration in the first stage to ensure the sensing accuracy, which results in more remaining duration in the second stage to perform energy beamforming. When $\Gamma$ further increases, the minimum harvested power is observed to decrease, as the estimated parameters of the ERs become less accurate in this case. This thus shows the trade-off between sensing accuracy and energy transmission, as shown in Remark \ref{remark}. 
	
	In the following, we compare the performance of our proposed design  with the following benchmark schemes. Note that in our proposed design, we set $\Gamma$ as the optimized value to maximize the minimum harvested energy. 
	\begin{itemize}	
		\item \textbf {Perfect CSI: }By considering that the CSI of $\{\mathbf{h}_{k}\}$ is perfectly known \textit{a priori}, the whole transmission block is allocated to the second stage for energy transmission, in which the energy beamforming $\mathbf{R}_{x}^{\star}$ is obtained by solving problem  (P2).
		Then the minimum harvested power among all the ERs is $\mathop{\min}\limits_{k\in\mathcal K} \left\{\boldsymbol{h}_{1}^{H} \mathbf{R}_{x}^{\star} \boldsymbol{h}_{1}, \ldots, \boldsymbol{h}_{K}^{H} \mathbf{R}_{x}^{\star} \boldsymbol{h}_{K} \right\}$.
		\item \textbf {Isotropic transmission: }The whole transmission block is used for energy transmission. In particular, the AP uses the identity transmit covariance matrix $\mathbf{R}_{x}=\frac{P_{\text{max}}}{N_{t}}\mathbf{I}$. 
		\item \textbf {Equal time allocation: }The whole transmission block is divided into two stages with equal durations for radar sensing and energy transmission, in which the sensing and energy beamformers are designed based on those in Section III. 
	\end{itemize}
	
\begin{figure}
		\setlength{\abovecaptionskip}{-10pt}
		\setlength{\belowcaptionskip}{-15pt}
		\centering
			\centering
			\includegraphics[width= 0.37\textwidth]{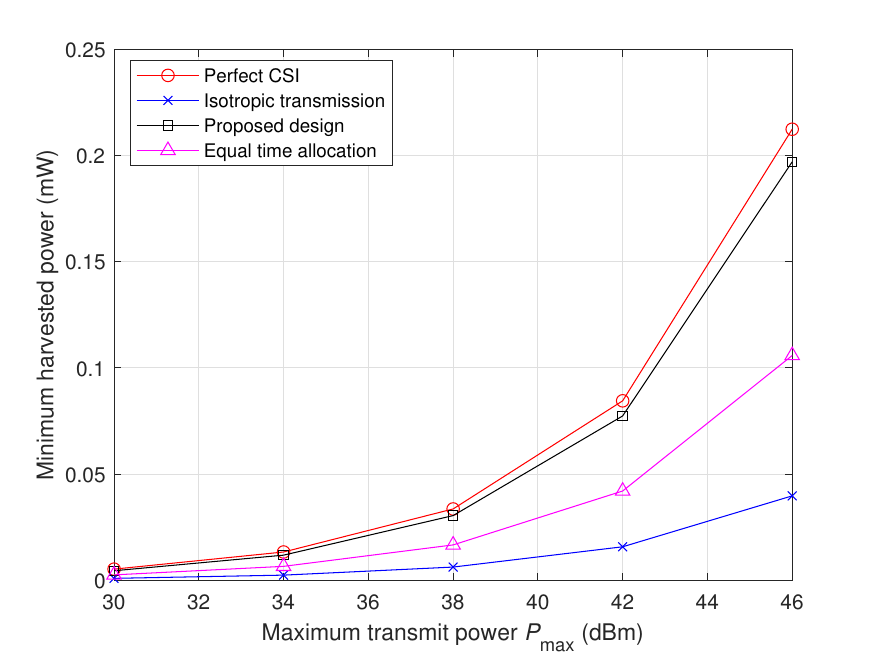}
		\DeclareGraphicsExtensions.
		\captionsetup{font={scriptsize},skip=1pt}
		\caption{The minimum harvested power versus the maximum transmit power $P_{\text{max}}$.}
		\label{figure4}
	\end{figure}

	Fig. \ref{figure4} shows the minimum average harvested power versus the maximum transmit power, where the numbers of transmit and receive antennas at the AP are set as $N_{t} = N_{r} = 8$. It is observed that our proposed design performs close to the  performance upper bound with perfect CSI. This shows the effectiveness of our proposal. Furthermore, the proposed design is observed to outperform the isotropic transmission, thanks to the transmit energy beamforming design for steering energy towards desired directions. Additionally, the proposed design is observed to outperform the scheme with equal time allocation, as the transmission energy in the benchmark is wasted  due to the lack of time allocation design. 
	
\begin{figure}
		\setlength{\abovecaptionskip}{-10pt}
		\setlength{\belowcaptionskip}{-15pt}
		\centering
			\centering
			\includegraphics[width= 0.37\textwidth]{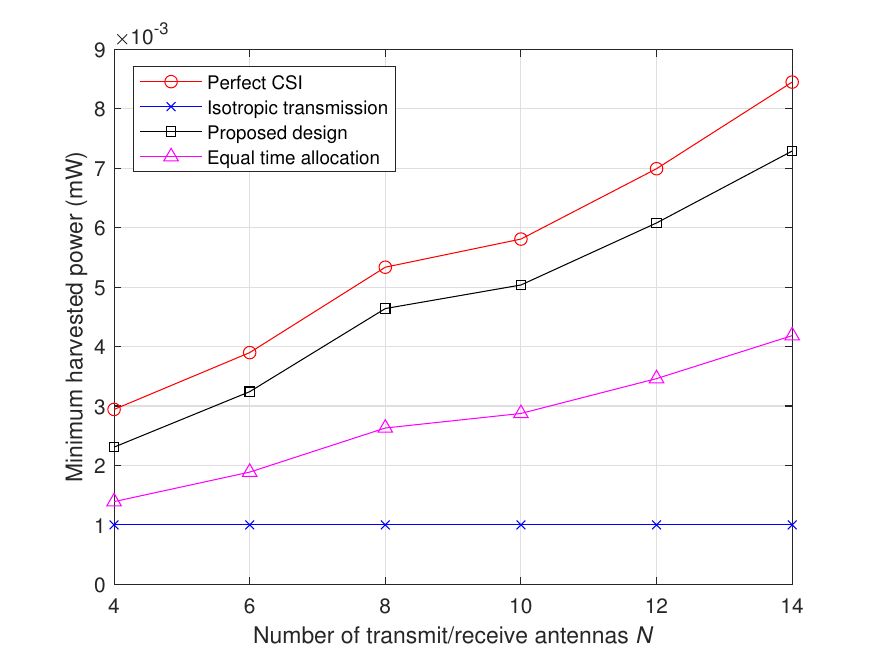}
		\DeclareGraphicsExtensions.
		\captionsetup{font={scriptsize},skip=1pt}
		\caption{The minimum harvested power versus number of antennas $N$ at the AP, where $N_{t}=N_{r}=N$.}
		\label{figure5}
	\end{figure}

	Fig. \ref{figure5} shows the performance versus the number of antennas at the AP, where the numbers of transmit antennas and receive antennas are the same (i.e., $N_{t} = N_{r} = N$),   and the maximum transmit power is set to be $P_{\text{max}}=30$ dBm. It is observed that the performance achieved by the proposed design, the scheme with perfect CSI, and the equal time allocation improves as $N$ increases, while  the performance of isotropic transmission remains unchanged. This is because in the first three schemes, energy beamforming can be steered towards the ERs. Furthermore, the proposed design is observed to significantly outperform the benchmark schemes with the isotropic transmission and the equal time allocation, which can be similarly explained as for Fig. \ref{figure4}.  
	\vspace{-0.1cm}
	\section{Conclusion}
	\vspace{-0.1cm}
	In this paper, we proposed a sensing-assisted energy beamforming approach to  wirelessly charge multiple ERs without requiring channel training. We presented a two-stage protocol, based on which the AP performs sensing to estimate the angles and path gains of the ERs for constructing the CSI in the first radar sensing stage, and then implements the energy beamforming based on the constructed CSI in the second energy transmission stage. We jointly designed the sensing duration and the joint beamforming to maximize the minimum harvested energy among all ERs, while ensuring a predetermined sensing accuracy requirement in the first stage. In the proposed design, the sensing accuracy is properly controlled to balance the time duration allocation between the two stages to optimize the ultimate energy harvesting performance. Numerical results showed that the proposed design performs close to the performance upper bound with perfect CSI, and outperforms other benchmark schemes.

\end{document}